\documentclass{article}
\usepackage{spconf,amsmath,graphicx}
\usepackage{mathrsfs}
\usepackage{subcaption}

\usepackage{blindtext}
\title{Speaker-invariant Affective Representation Learning \\via Adversarial Training}

\name{Haoqi Li$^{\star}$ \qquad Ming Tu$^{\dagger}$ \qquad Jing Huang$^{\dagger}$ \qquad Shrikanth Narayanan$^{\star}$ \qquad Panayiotis Georgiou$^{\star}$}

\address{$^{\star}$Dept. of Electrical and Computer Engineering, University of Southern California, USA \\
  $^{\dagger}$JD AI Research, Mountain View, USA\\}
  
\usepackage{color}
\definecolor{mypink1}{rgb}{0.858, 0.188, 0.478}
\definecolor{mycolor3}{rgb}{0, 0.5, 0.5}
\definecolor{mycolor2}{rgb}{0, 0, 0.9}
\definecolor{mygreen}{rgb}{0, 1, 0}
\definecolor{mymaroon}{rgb}{0.5, 0, 0}

\newcommand{\Pg}[1]{\noindent\textbf{\color{black}#1}}

\begin{document}
\ninept
\maketitle
\begin{abstract}

Representation learning for speech emotion recognition is challenging due to labeled data sparsity issue and lack of gold-standard references. In addition, there is much variability from input speech signals, human subjective perception of the signals and emotion label ambiguity.  
In this paper, we propose a machine learning framework to obtain speech emotion representations by limiting the effect of speaker variability in the speech signals.
Specifically we propose to disentangle the speaker characteristics from emotion through an adversarial training network in order to better represent emotion. 
Our method combines the gradient reversal technique with an entropy loss function to remove such speaker information.
Our approach is evaluated on both IEMOCAP and CMU-MOSEI datasets. 
We show that our method improves speech emotion classification and increases generalization to unseen speakers.

\end{abstract}
\vspace{-1.8ex}
\section{Introduction}
\label{sec:intro}
\vspace{-0.3ex}
Human speech signals contain rich linguistic and paralinguistic information. Linguistic information is encoded at different temporal scales ranging from phoneme to sentence and discourse levels. More importantly, speech signal encodes speaker characteristics and affective information.
All information above is jointly modulated and intertwined in the human-produced speech acoustics and it is difficult to dissociate these various components simply from features, such as those from the time waveform or its transformed representations e.g., Mel filterbank energies. 

Representation learning of speech \cite{chung2019unsupervised, pascual2019learning, li2017unsupervised}, i.e., the transformation from low-level acoustic descriptors to higher-level representations, has received significant attention recently. 
Traditional methods focus on using supervised learning, specifically multi-task learning \cite{caruana1997multitask} to extract specialized representations of particular targets.
However, target representations are easily contaminated by undesired factors, such as noise, channel or source (speaker) variability. These are difficult to eliminate due to the complexity and entanglement of information sources in the speech signal.

Emotion recognition systems are further greatly affected by source variability, be that speaker, ambient acoustic conditions, language, or socio-cultural context \cite{schuller2018speech}.
Limited domain data and labeling costs have resulted in many systems that are only evaluated within domain and are not robust to such variability. For example mismatch between training and evaluation sets, such as speaker variations \cite{zhan1997speaker} and domain condition incongruity \cite{ganin2016domain}, make it challenging to obtain robust emotion representations across different speakers and domains.

In this work, we propose an adversarial training framework to learn robust speech emotion representations. 
We specifically aim to remove speaker information from the representation, as that is one of the most challenging and confounding issues in speech emotion recognition (SER). 
Note that many SER systems have addressed this issue through normalization of features, but these ad-hoc solutions lack generalization within complex learning frameworks \cite{sethu2007speaker, busso2011iterative}.

In our work, inspired by the domain adversarial training (DAT) \cite{ganin2016domain}, we propose a neural network model and an adversarial training framework with an entropy-based speaker loss function to relieve speaker variability influences. 
Considering the adversarial training strategy and entropy-based objective function, we name our model \emph{Max-Entropy Adversarial Network} (\emph{MEnAN}). 
We demonstrate the effectiveness of the proposed framework in SER within- and across-corpora.
We show that MEnAN can successfully remove the speaker-information from extracted emotion representations, and this disentanglement can further effectively improve speech emotion classification on both the IEMOCAP and CMU-MOSEI datasets.

\vspace{-1.8ex}
\section{Related work}
\label{sec:related_work}
\vspace{-0.3ex}

Robust representations of emotions in speech signals have been investigated via pre-trained denoising autoencoders \cite{ghosh2016representation}, end-to-end learning from raw audio \cite{trigeorgis2016adieu}, unsupervised contextual learning \cite{li2017unsupervised}, and multi-task learning \cite{zhang2017cross} etc. In a different way, we apply GANs based adversarial training to generate robust representations across domains (speaker to be specific) for speech emotion recognition.
Among previous work on SER, GANs are mainly utilized to learn discriminative representations \cite{chang2017learning} and conduct data augmentation \cite{sahu2018enhancing}. Our method is different in that we aim to disentangle speaker information and learn speaker-invariant representations for SER.

Recently, within speech applications, domain adversarial training (DAT) techniques have been applied on cross-corpus speaker recognition \cite{wang2018unsupervised}, automatic speech recognition \cite{shinohara2016adversarial,sun2017unsupervised, meng2018speaker} and SER \cite{abdelwahab2018domain, tu2019towards} to deal with the domain mismatch problems. Compared to the two most related studies \cite{abdelwahab2018domain, tu2019towards}, our proposed MEnAN is different from DAT: 1) we argue that simple gradient reversal layer in DAT may not guarantee domain-invariant representation: simply flipping the domain labels can also fool the domain classifier however the learned representation is not necessary to be domain invariant. 2) we propose a new entropy-based loss function for domain classifier to induce representations that maximize the entropy of the domain classifier output, and we show the learned representation is better than DAT for speech emotion recognition.

\vspace{-1.8ex}
\section{Methodology}
\label{sec:method}
\vspace{-0.3ex}
Our goal is to obtain an embedding from a given speech utterance, in which emotion-related information is maximized while minimizing the information relevant to speaker identities. This is achieved by our proposed adversarial training procedure with designed loss functions which will be introduced in this section.

\subsection{Model structure}
\label{subsec:model_structure}
\begin{figure}[htb]
\centering
\includegraphics[width=8.6cm]{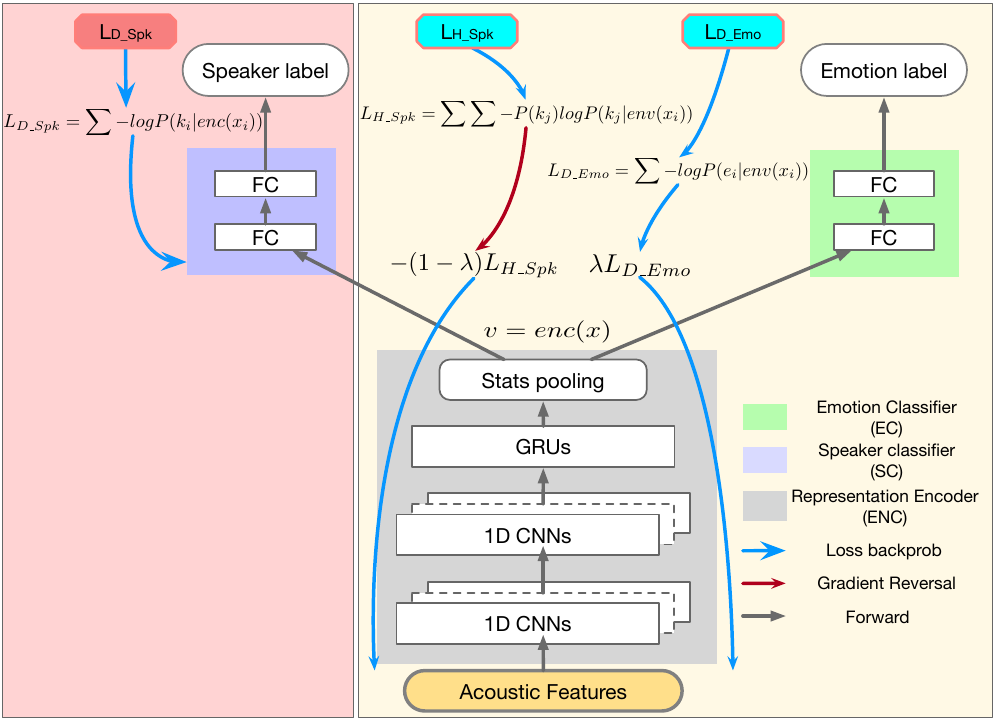}
\caption{Model structure with loss propagating flow}
\label{fig:model}
\end{figure}

Our proposed model is built based on a multi-task setting with three modules: the representation encoder (ENC), the emotion category classification  module (EC) and the speaker identity classification module (SC). The structure of our model is illustrated in Figure \ref{fig:model}.

The ENC module has three components: (1) stacked 1D convolutional layers; (2) recurrent neural layers and (3) statistical pooling layers. 
The sequence of acoustic spectral features is first input to multiple 1D CNN layers. The CNN kernel filters shift along the temporal axis and include the entire spectrum information per scan, which is proven to have better performance than other kernel structure settings by \cite{huang2017characterizing}. 
CNN filters with different weights are utilized to extract different information from same input features and followed by recurrent layers to capture context and dynamic information within each speech segment.
Then, we add the statistical pooling functions, including maximum, mean and standard deviation in our model, to map a variable speech segment into an embedding with a fixed dimension. 

This fixed dimension representation embedding, as the output of ENC, is further connected with two sub-task modules: the emotion classifier (EC) and speaker classifier (SC), which are both built with stacked fully connected layers.

With normal training settings, our model can be regard as a multi-task learning framework. Moreover, our model can be regarded as a speech emotion recognition system if we only keep the EC and ENC components.

\subsection{Difference with prior work}

In domain adversarial training \cite{ganin2016domain, abdelwahab2018domain,tu2019towards}, one gradient reversal layer is usually added to the domain classifier (SC in our case) to reverse the gradient flow in order to generate the domain-invariant (speaker-invariant) features.
The usage of the gradient reversal layer ensures the (desired) lower performance of domain classifier (SC in our case), however, it often fails to guarantee the domain (speaker) information has been fully removed \cite{liu2019transferable}. For instance, in this approach, it is highly likely that even a lower performance of SC will only map a particular speaker to other target speakers with similar sounds instead of properly removing the speaker identity information, likely picking up the second-best speaker match.

Our proposed training method is different from the existing strategy as it attempts to completely remove all speaker information.

\subsection{Emotion representation adversarial training}
We now describe the adversarial training strategy and the designed loss function in detail. 
Our training dataset $\mathcal{D}=\{(x_1,e_1,k_1),...,(x_N,e_N,k_N)\}$ contains $N$ pairs of $(x_i,e_i,k_i)\in(\mathcal{X},\mathcal{E},\mathcal{K})$, in which speech segment $x_i$ is produced by the speaker $k_i$ with emotion $e_i$. $\mathcal{X}$, $\mathcal{E}$ and $\mathcal{K}$ are the sets of whole speech utterances, emotion labels and all speakers respectively.

Our training strategy is similar to generative adversarial networks (GANs) \cite{goodfellow2014generative}. 
The system has two output paths. On one path (left Fig.\ref{fig:model}), we attempt to accurately estimate the speaker information (loss $L_{D\_Spk}$). On the other path (right Fig.\ref{fig:model}), we attempt to estimate the emotion label (loss $L_{D\_Emo}$) and remove speaker information (loss $L_{H\_Spk}$). 
Both estimators (SC and EC) employ the same representation encoder (ENC) but that is only updated from the right-side loss back propagation.

The output of ENC, denoted as $v$, is the speaker-invariant emotion representation we try to obtain. We have $v=enc(x)$.

\subsubsection{Training of SC}
\label{subsubsec:spk_classifier}
The speaker classifier (SC) can be regarded as a discriminator which is trained to distinguish speaker identities based on a given encoder output $v$ and has no influence in the training of $v$.
The SC is trained by minimizing $L_{D\_Spk}$, the cross entropy function as in (\ref{eq:ce_spk}):
\begin{equation}\label{eq:ce_spk}
L_{D\_Spk} = 
\sum_{(x_i,e_i,k_i)\in \mathcal{D}}-logP(k_i|enc(x_i))
\end{equation}
In this training step, weights of ENC and EC are frozen. 
Only parameters of SC are optimized to achieve higher speaker classification accuracy from a given representation $v$. 

\subsubsection{Training of ENC and EC}
\label{subsubsec:ebd_classifier}
Under adversarial training we need to ensure the ENC output contains emotion-related information, while it is also optimized to confuse and make it difficult for SC to distinguish speaker identities.

Thus, we need to optimize ENC to increase the uncertainty or randomness of SC's outputs. Mathematically, we want to maximize the entropy value of SC's output. The entropy of SC's output, denoted as $L_{H\_Spk}$, is defined as 
\begin{equation}
L_{H\_Spk} =  \sum_{(x_i,e_i,k_i)\in \mathcal{D}}\sum_{k_j\in\mathcal{K}}- P(k_j|env(x_i))logP(k_j|env(x_i)) 
\end{equation}

Maximizing entropy would promote equal likelihood for all speakers:
\begin{equation}
P(k_j|env(x_i)) = P(k_q|env(x_i)) \quad \forall k_j,k_q \in \mathcal{K} 
\end{equation}
This differs, as mentioned above, from simply picking up a different speaker, since that may lead into selecting a ``second-best" similar sounding speaker. Our proposed loss function removes all (correct or wrong) speaker information.

In addition, the performance of emotion classifier is optimized by minimizing the cross entropy loss $L_{D\_Emo}$ from EC's output:
\begin{equation}
L_{D\_Emo} =  \sum_{(x_i,e_i,k_i)\in \mathcal{D}} -logP(e_i|env(x_i))
\end{equation}
To combine these two objective functions above together, we flip the sign of $L_{H\_Spk}$ to do a gradient reversal and minimize the weighted overall loss sums. The final objective loss function is written as:
\begin{equation}\label{eq:gan_loss}
L(\theta_{ENC}, \theta_{EC}) = \lambda L_{D\_Emo} - (1-\lambda)L_{H\_Spk}
\end{equation}
where $ \lambda \in (0,1)$ is a parameter adjusting the weighting between two types of loss functions. 

In this training step, weights of SC are frozen. Only parameters of ENC and EC  are  optimized. 
Modules with corresponding loss back propagation flows are shown in Fig.\ref{fig:model}. With this iterative training scheme, we expect the proposed model can ultimately relieve the impact of speaker variability thus improve the SER performance.

\vspace{-0.2em}
\section{Dateset}
\label{sec:data}
\vspace{-0.2em}
Two datasets are employed to evaluate the proposed MEnAN based emotion representation learning in our work:

The \Pg{IEMOCAP} dataset \cite{busso2008iemocap} consists of five sessions of speech segments with categorical emotion annotation, and there are two different speakers (one female and one male) in each session. 
In our work, we use both improvised and scripted speech recordings and merge \textit{excitement} with \textit{happy} to achieve a more balanced label distribution, a common experiment setting in many studies such as \cite{ghosh2016representation, neumann2019improving,gideon2017progressive}. Finally, we obtain 5,531 utterances selected from four emotion classes (1,103 angry, 1,636 happy, 1,708 neutral and 1,084 sad). 

The \Pg{CMU-MOSEI} dataset \cite{zadeh2018multimodal} contains 23,453 single-speaker video segments carefully chosen from YouTube. This database includes 1000 distinct speakers, and are gender balanced with an average length of 7.28 seconds. 
Each sample has been manually annotated with a [0,3] Likert scale on the six basic emotion categories: happiness, sadness, anger, fear, disgust, and surprise.
The original ratings are also binarized for emotion classification: for each emotion, if a rating is greater than zero, it is considered that there is presence of that emotion, while a zero results in a false presence of that emotion. Thus, each segment can have multiple emotion presence labels.

IEMOCAP provides a relatively large number of samples within each combination across different speakers and emotions, making it feasible to train our speaker-invariant emotion representation. 
We mainly use CMU-MOSEI for evaluation purposes, since it includes variable speaker identities, and to establish cross-domain transferability of MEnAN. 

\vspace{-1.2em}
\section{Experiment Setup}
\label{sec:exp_setup}
\vspace{-0.2em}
\Pg{Feature extraction: } 
In this work we utilize 40 dimensional Log-Mel Filterbank energies (Log-MFBs), pitch and energy. All these features are extracted with a 40 ms analysis window with a shift of 10 ms.
For pitch, we employ the extraction method in \cite{ghahremani2014pitch}, in which the normalized cross correlation function (NCCF) and pitch (f0) are included for each frame. 
We do not perform any per-speaker/sample normalization.

\Pg{Data augmentation:} 
To enrich the dataset, we perform  data augmentation on IEMOCAP. Similar to \cite{tang2018end}, we create multiple data samples for training by slightly modifying the speaking rate with four different speed ratios, namely 0.8, 0.9, 1.1 and 1.2. 

\Pg{General settings:}
To obtain a reliable evaluation of our model, we need to ensure unseen speakers for both validation and testing. Thus, we conduct 10 fold leave-one-speaker out cross-validation scheme.
More specifically, we use 8 speakers as the training set, and for the remaining session (two speakers), we select one speaker for validation and one for testing. 
We then repeat the experiment with the two speakers switched.

In addition, considering the variable length of utterances, we only extract the middle 14s to calculate acoustic features for utterance whose duration is longer than 14s (2.07\% of the total dataset) \cite{dai2019learning}, since important dynamic emotional information is usually located in the middle part and lengthy inputs would have negative effect on emotion prediction \cite{huang2018speech}. 
For utterances shorter than 14s, we use the cycle repeat mode \cite{huang2018speech} to repeat the whole utterance to the target duration. The idea of this cycle repeat mode is to make emotional dynamic cyclic and longer, which facilitates the training process of utterances of variable duration.
\vspace{-0.3em}
\subsection{Model configurations}
\label{subsec:model_config}
The detailed model parameters and training configurations are shown in Table \ref{tab:config}.

\begin{table}[ht]
\centering
\scalebox{0.75}{
\begin{tabular}{c|l}
\hline
\begin{tabular}[c]{@{}c@{}}Training \\ details:\end{tabular} & \begin{tabular}[c]{@{}l@{}}Adam optimizer (lr=0.001) +  polynomial learning rate decay ; \\ batch size=16; epochs=300; $\lambda$=0.5 \end{tabular}                                                                                                                                                                               \\ \hline
ENC                                                          & \begin{tabular}[c]{@{}l@{}}Conv1D(in\_ch=43,out\_ch=32, kernel size=10, stride=2, padding=0), PReLU\\ Conv1D(in\_ch=32,out\_ch=32, kernel size=5,stride=2, padding=0), PReLU\\ GRU(in\_size=32, hidden\_size=32, num\_layers=1)\\ Linear(in=32, out=32), PReLU\\ Statistical Pooling[Mean, Std, Max]\end{tabular} \\ \hline
EC                                                           & \begin{tabular}[c]{@{}l@{}}Linear(in=32*3, out=32) PReLU\\ Linear(in=32, out=10) PReLU\\ Linear(in=10, out=4 )\end{tabular}                                                                                                                                                                                       \\ \hline
SC                                                           & \begin{tabular}[c]{@{}l@{}}Linear(in=32*3, out=32) PReLU\\ Linear(in=32, out=10) PReLU\\ Linear(in=10, out=8)\end{tabular}                                                                                                                                                                                        \\ \hline
\end{tabular}
}
\vspace{-0.8em}
\caption{Model structure and training configuration details}
\vspace{-0.8em}
\label{tab:config}
\end{table}

\vspace{-1.1em}
\section{Results and Discussion}
\label{sec:results}
\Pg{Evaluation on IEMOCAP}

For comparison purposes, we also train the EC only model, multi-task learning model and DAT model \cite{tu2019towards} with regular cross entropy loss under the same configuration. Both the weighted accuracy (WA, the number of the correctly classified samples divided by the total number of samples) and the unweighted accuracy (UA, the mean value of the recall for each class) are reported.
The Table \ref{tab:iemocap_results} shows the emotion classification accuracy (\%) on both validation and testing, and we also include their differences ($\Delta$).

\begin{table}[ht]
\scalebox{0.85}{
\begin{tabular}{cccclccc}
\hline

                                  & \multicolumn{3}{c}{WA}         &  & \multicolumn{3}{c}{UA}         \\ \cline{2-8}
                 & Val   & Test           & $\Delta$    &  & Val   & Test           & $\Delta$     \\ \hline
EC only model    & 58.50 & 55.92          & -2.56                                                         &  & 59.94 & 57.45          & -2.49                                                         \\ 
Multi-task model & 59.24 & 55.90          & -3.34                                                         &  & 60.52 & 57.28          & -3.24                                                         \\ 
DAT model & 58.28 & 56.68                 & -1.60                                                         &  & 60.16 & 58.48          & -1.68                                                         \\ 
Proposed MEnAN        & 58.85 & \textbf{58.62} & -0.23                                                         &  & 60.24 & \textbf{59.91} & -0.33                                                         \\ \hline
\end{tabular}
}
\vspace{-0.8em}
\caption{Classification accuracy (\%) comparison on IEMOCAP}
\label{tab:iemocap_results}
\end{table}
First, we observe that our model achieves the best classification accuracy in the test case among all models. 
To the best of our knowledge, the best results from the literature on IEMOCAP with similar settings are generally around 60\% \cite{xia2015leveraging, neumann2019improving}. 
We achieve a UA of 59.91\% which is comparable with the state of the art results. 
However, strict comparisons remain difficult because there are no standardized data selection rules or train/test splits. 
For example, some did not use speaker independent split \cite{neumann2019improving} or only used improvised utterances. Some did not clearly specify which speaker in each session was used for validation and testing respectively \cite{lee2015high} or performed per-speaker normalization in advance \cite{busso2011iterative, sahu2019multi}. 

Second, we notice that there is a large difference of $\Delta$ value among all four models.
Compared with others, we find that the multi-task learning model can achieve a better performance on the validation set.
However, the extra gain from speaker information can also lead a significant mismatch during the evaluation of unseen speakers, as indicated by the large value of $\Delta$. 
Compared with DAT model, our MEnAN model gains better classification accuracy with smaller $\Delta$.
This supports our claim of the MEnAN's advantage over DAT. 
The small $\Delta$ in our model suggests our embedding has better generalization ability and is more robust to unseen speakers.
To illustrate this, we plot t-SNE of emotion representation, i.e., $enc(x)$, on two unseen speakers. 

\begin{figure}[ht]
\vspace{-0.3em}
\begin{minipage}[b]{1.0\linewidth}
  \centering
  \centerline{\includegraphics[width=7.5cm]{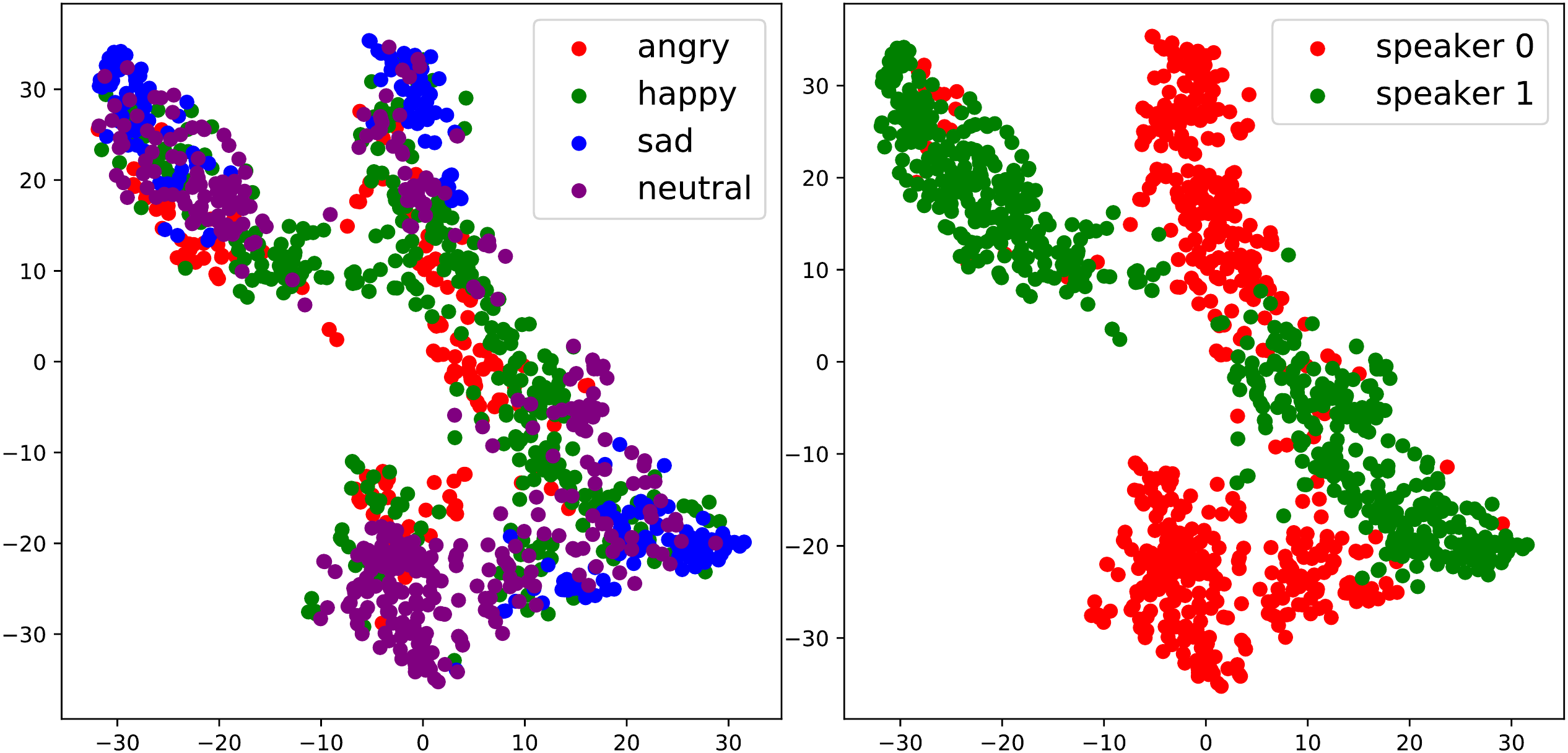}}
  \centerline{(a) Multi-task learning model}\medskip
\end{minipage}
\begin{minipage}[b]{1.0\linewidth}
  \centering
  \centerline{\includegraphics[width=7.5cm]{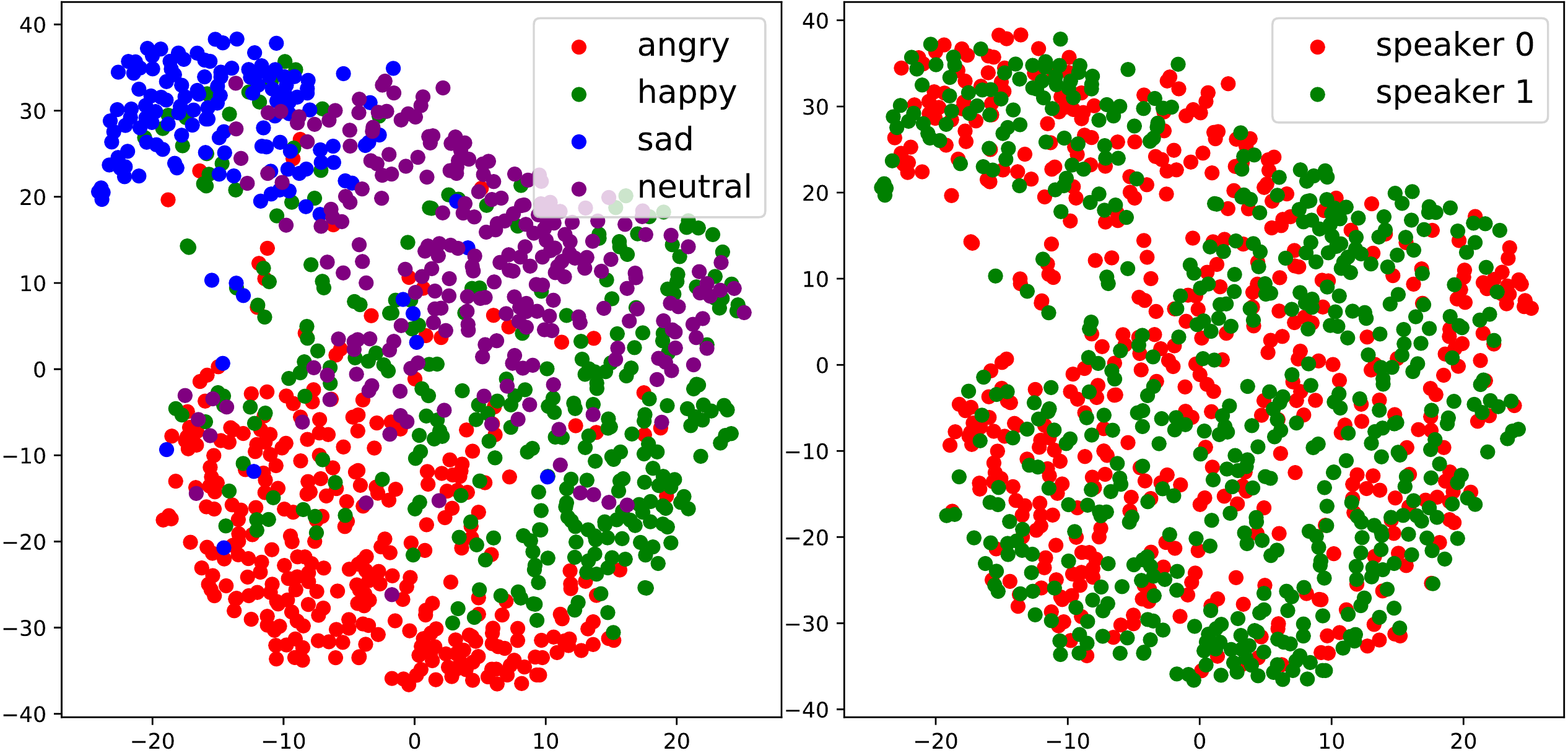}}
  \centerline{(b) MEnAN model}\medskip
  \vspace{-1em}
\end{minipage}
\vspace{-1em}
\caption{t-SNE plot of emotion embedding with both 4 emotion labels (left) and 2 speaker labels (right) for multi-task model and our proposed MEnAN model.}
\label{fig:tsne}
\vspace{-1em}
\end{figure}

As shown in Fig.\ref{fig:tsne}, in the multi-task learning setting, it is obvious that the speaker's characteristics and emotion information are entangled with each other, which makes this representation less generic on unseen speakers. 
For our proposed MEnAN, the speaker representations of different speakers on the 2D space are well mixed and independent of speaker labels; while different emotion segments are more separable in the embedding manifold. 
These results further demonstrates the effectiveness and robustness of proposed model.

\Pg{Evaluation on CMU-MOSEI}

In addition, we test our system on the CMU-MOSEI dataset (cross-corpus setting).
As mentioned before, the CMU-MOSEI has a large variability in speaker identities, which is a suitable corpus to evaluate our model's generalization ability on unseen speakers.
It also introduces a challenge stemming from the different annotation methodology and the inherent effect on the interpretation of labels.

To match emotion labels of IEMOCAP, we only consider samples with positive ratings in the categories of ``happiness", ``sadness" and ``anger". 
Samples with zero ratings of all six emotion categories are also included with the label ``Neutral". 
Finally, 22,323 samples are selected and four-class emotion classification evaluation are performed. The prediction is considered to be correct if the rating of that predicted emotion originally has a positive value.
In Table \ref{tab:cmumosei}, we report the mean, minimum and maximum of the classification accuracy (\%) evaluated on the pretrained model of each fold from the 10-fold cross validation of IEMOCAP. 

\begin{table}[htp]
\centering
\begin{tabular}{cccc}
\hline
          & mean  & min & max \\ \hline
EC only model   & 31.35 & 27.14    &  34.96    \\ 
DAT model   & 32.34 & 25.91    &  37.85    \\ 
Proposed MEnAN & \textbf{33.24} & \textbf{28.84}    &  \textbf{39.85}   \\ \hline
\end{tabular}
\vspace{-0.5em}
\caption{Emotion classification accuracy (\%) on CMU-MOSEI}
\vspace{-1.5em}
\label{tab:cmumosei}
\end{table}

We observe that MEnAN model has the best performance among all three models, and it achieves better classification accuracy with 1.89\% improvement on the mean value and with 4.89\% on the best model compared with the EC only model. Considering that all speakers of these evaluation samples are not seen during the training, these results suggest our adversarial training framework can provide more robust emotion representation with better speaker-invariant property and achieve improved  performance in the emotion recognition task. 
\vspace{-0.3em}
\section{Conclusion}
\label{sec:conclusion}
\vspace{-0.2em}
Compared with other representation learning tasks, the extraction of speech emotion representation is challenging considering the complex hierarchical information structures within the speech, as well as the practical low-resource (labeled) data issue. 
In our work, we use an adversarial training strategy to generate speech emotion representations while being robust to unseen speakers.
Our proposed framework MEnAN, however, is not limited to the emotion recognition task, and it can be easily applied to other domains with similar settings e.g., cross-lingual speaker recognition. 
For further work, we plan to combine the domain adaption techniques with our proposed model to employ training samples from different corpora. 
For example, we can utilize speech utterances from speaker verification tasks to obtain more robust speaker information.

\vspace{-0.3em}

\bibliographystyle{IEEEbib}
\bibliography{strings,refs}

\end{document}